\documentclass{article}

\usepackage{titling}
\usepackage{graphicx,color,psfrag}
\usepackage{fancyheadings}
\usepackage{arxiv}

\usepackage[backend=biber,style=authoryear-ibid,maxcitenames=2,maxbibnames=20,firstinits=true]{biblatex}
\bibliography{plwh,Books,data}
\AtEveryBibitem{\clearfield{doi}}
\AtEveryBibitem{\clearfield{isbn}}
\AtEveryBibitem{\clearfield{issn}}

\DeclareNameAlias{sortname}{family-given}
\renewbibmacro{in:}{}
\DeclareFieldFormat[article, inbook, incollection, inproceedings, misc, thesis, unpublished]{title}{#1}
\DeclareFieldFormat[article]{volume}{{#1}} 
\DeclareFieldFormat[article, inbook, incollection, inproceedings, misc, thesis, unpublished]{number}{\mkbibparens{#1}} 
\renewbibmacro*{volume+number+eid}{
  \printfield{volume}
  \printfield{number}
}

\title{Predicting Chronic Homelessness: The Importance of Comparing Algorithms using Client Histories}
\author{Geoffrey Messier  \\
  University of Calgary  \\
  2500 University Dr.~NW, Calgary, AB, Canada, T2N 1N4\\
  gmessier@ucalgary.ca
  \and
  Caleb John\\
  University of Calgary  \\
  2500 University Dr.~NW, Calgary, AB, Canada, T2N 1N4\\
  ctjohn@ucalgary.ca
  \and
  Ayush Malik\\
  University of Calgary  \\
  2500 University Dr.~NW, Calgary, AB, Canada, T2N 1N4\\
  ayushmalik118@gmail.com
}

\date{}

\newcommand{\bE}{\begin{enumerate}}
\newcommand{\eE}{\end{enumerate}}
\newcommand{\bI}{\begin{itemize}}
\newcommand{\eI}{\end{itemize}}

\begin{document}

\maketitle

\begin{abstract}
This paper investigates how to best compare algorithms for predicting chronic homelessness for the purpose of identifying good candidates for housing programs.  Predictive methods can rapidly refer potentially chronic shelter users to housing but also sometimes incorrectly identify individuals who will not become chronic (false positives).  We use shelter access histories to demonstrate that these false positives are often still good candidates for housing.  Using this approach, we compare a simple threshold method for predicting chronic homelessness to the more complex logistic regression and neural network algorithms.  While traditional binary classification performance metrics show that the machine learning algorithms perform better than the threshold technique, an examination of the shelter access histories of the cohorts identified by the three algorithms show that they select groups with very similar characteristics.  This has important implications for resource constrained not-for-profit organizations since the threshold technique can be implemented using much simpler information technology infrastructure than the machine learning algorithms.
\end{abstract}

\section{Introduction}
\label{sec:intro}

Data science has the potential to be a powerful tool for helping to end homelessness.  To date, there has been considerable activity aimed at using data analysis and machine learning algorithms to assess the risk that an individual will experience a homelessness related event.  Most algorithms applied to homelessness problems are binary classifiers that seek to determine whether an event will or will not occur.  Evaluating the risk of entering or re-entering homelessness has been well explored \parencite{diguiseppi-gt-2020, chan-h-2017, kube-a-2019, greer-al-2016, hong-b-2018, shinn-m-2013, gao-y-2017, byrne-t-2019}.   These studies evaluate their algorithms using {\em classification metrics} such as true positive rate  (sensitivity), false positive rate (specificity), false negative rate (miss rate) and true negative rate (selectivity) \parencite{powers-dmw-2008}.  

Housing First programming is another important aspect of addressing homelessness.  The goal of Housing First is to quickly connect people experiencing homelessness with permanent housing without pre-conditions \parencite{goering-p-2014, usich-2015}.  It is common to prioritize people experiencing chronic homelessness for Housing First programs \parencite{gaetz-s-2014, aubry-t-2015, toros-h-2018}.  These individuals can be identified using one of several chronic homelessness definitions \parencite{byrne-t-2015, di-community-report-2019, synder-sg-2008, employment-social-development-2019}.  However, these definitions require a minimum of 6-12 months of homelessness before identifying an individual as chronic.

Clearly, forcing a new entrant to homelessness to accumulate many months of shelter use before being prioritized for housing support is undesirable.  Predictive data analysis algorithms have the potential to identify good candidates for Housing First programs much more quickly \parencite{vanberlo-b-2020, toros-h-2018}.  These authors demonstrate the efficacy of their methods using classification metrics with \parencite{toros-h-2018} going a step further to also estimate system cost savings.  Both papers demonstrate good results but the complexity of their methods are beyond the reach of the information technology (IT) capabilities of many resource constrained not-for-profit organizations.

The contribution of this paper is to demonstrate the importance of going further than classification metrics when evaluating the best method to identify individuals at risk of chronic homelessness.   In particular, our approach will reveal that it is not necessary to use a high complexity machine learning algorithm to select a good cohort to prioritize for housing services.  Using classification metrics alone can be misleading due to how chronic homelessness is defined.  Unlike a first entry into homelessness, chronic homelessness is not a single observable event.  Rather, it is a pattern of events that evolve over time to eventually satisfy a definition.  When a classifier accidentally selects a person who does not meet the chosen definition for chronic homelessness, it is penalized by an increase in its false positive rate.  However, the false positive individual may still be an excellent candidate for housing support.

For example, consider two algorithms attempting to predict whether an individual will eventually meet the chronic definition criterion of 180 days of homelessness within a year \parencite{employment-social-development-2019}.  The first algorithm selects a non-chronic individual with 179~days of homelessness and the second a non-chronic individual with only 1~day of homelessness.  Both selections would result in the same increase in false positive rate but the first algorithm has clearly done a better job of choosing a person in need.

To illustrate our point, we will compare three different binary classifiers for predicting chronic homelessness: logistic regression, a neural network and a simple shelter stay threshold test.  While we will evaluate them using classification metrics, we will also perform a statistical analysis of the shelter access histories of the cohorts identified by the algorithms.  These cohorts contain individuals who do eventually experience chronic homelessness (the true positives) and those who do not (the false positives).  Comparing the shelter access histories of the cohorts identified for support reveals that the simple threshold test provides comparable and, in some ways, better performance than the more complex algorithms.

This has important implications.  Implementing a sophisticated machine learning triage tool, like a neural network,  for front-line use would require a shelter to make a considerable investment in IT infrastructure and expertise.  This investment is unnecessary if a simple threshold test implemented in a spreadsheet by non-technical staff could achieve the same result.

While the focus of this paper is using emergency shelter client data to identify good candidates for housing, we should emphasize that we do not feel this is the only tool that should be used to triage clients for support programs \parencite{aubry-t-2015}.  Predictive data-based triage techniques should be incorporated into a housing and support program that includes counsellors and case workers who are reaching out to clients through a variety of mechanisms.  Data analysis adds value in this type of environment by helping to spot the ``under-the-radar'' clients who are not engaging with staff and may fall through the cracks, particularly in very busy shelters.   Any individual identified by the techniques discussed in this paper should not be immediately placed into housing but should instead be referred to a case manager for a comprehensive consultation.

The techniques we consider are evaluated using 13 years of anonymized client data from a major North American shelter.  The characteristics of this dataset, the pre-processing of the data and the data features used to predict chronic shelter use are described in Section~\ref{sec:data}.  The three predictive techniques are presented in Section~\ref{sec:alg} and their performance is compared using classification metrics and cohort shelter access histories in Section~\ref{sec:results}.  Concluding remarks are made in Section~\ref{sec:concl}.

\section{Description of Data}
\label{sec:data}

This is a secondary data analysis performed on anonymized shelter stay records collected at the Calgary Drop-In Centre (DI) between July 1, 2007 and January 20, 2020.  The data anonymization and client privacy protocol used during this study was approved by the University of Calgary Conjoint Faculties Research Ethics Board.  The dataset consists of 5,431,521 entries for 41,935 unique client profiles.  Each entry records a single timestamped interaction with a client that could be accessing sleep services, accessing counselling services, a security incident resulting in a ban from shelter or a log note.  To mitigate left censoring of the data, we exclude all clients who have their first sleep date appear prior to July 1, 2009 which removes a total of 19,967 clients.  To mitigate right censoring of the data, we also exclude all clients who have their first sleep date appear after January 20, 2018 which removes a total of 3,570 clients.  As a result, 18,398 individuals are retained in the dataset (43.9\% of the original dataset population).

The objective of this study is to use this dataset to train and test techniques for predicting whether a new shelter client will eventually become a chronic shelter user.  The Canadian federal definition of chronic homelessness is used to classify which individuals in the dataset develop into chronic shelter users.  This definition includes ``individuals who are currently experiencing homelessness and who meet at least 1 of the following criteria: they have a total of at least 6 months (180 days) of homelessness over the past year or they have recurrent experiences of homelessness over the past 3 years, with a cumulative duration of at least 18 months (546 days)'' \parencite{employment-social-development-2019}.  A total of 1,549 of the 18,398 clients (8.4\%) satisfy this definition and are designated as chronic in the dataset.

Many of the entries in the original DI dataset contain comments entered by DI staff related to their interaction with the client.  To maintain client privacy, these comment fields are stripped by DI IT staff as part of the anonymization process.  The DI IT staff replaced each comment field by a count of how many keywords occur in each comment.  The keywords are divided into the following categories:  Police, Emergency Medical Services (EMS), Health, Violence and Addiction.  For example, a common Police word was ``CPS'' (Calgary Police Services) and a common Violence word was ``fight''.  A comment field with one occurrence of ``CPS'' and two occurrences of ``fight'' would have a count of 1 for Police, a count of 2 for Violence and zero counts for all other keyword categories.

The techniques used in this paper to identify likely chronic shelter users will generate those predictions using data collected in a 90 day time window that starts from the first day a client sleeps in shelter.  This time window was chosen as a compromise between collecting enough data for a reliable prediction while still identifying individuals for housing supports as early in their shelter timeline as possible.  The data features generated for each client are summarized in Table~\ref{tb.features} including a column indicating whether the feature is a client's static characteristic or a count of the total number of each type of record over the 90 day period.  These features serve as the input to the prediction algorithms described in Section~\ref{sec:alg}.  

\begin{table}[htbp]
\centering
\begin{tabular}{c|c|c}
Field & Operation & Description \\ \hline
Age & Static & Client age. \\
Bar & Total Count & Records noting a ban from shelter services. \\
Sleep & Total Count & Sleep records.\\
Log & Total Count & General log entries.\\
Counsellor & Total Count & Counselling entries.\\
Police & Total Count & Records with one or more Police keywords. \\ 
EMS & Total Count &  Records with one or more EMS keywords. \\ 
Health & Total Count &  Records with one or more Health keywords. \\ 
Violence & Total Count &  Records with one or more Violence keywords. \\ 
Addiction & Total Count &  Records with one or more Addiction keywords. \\ 
\end{tabular}
\caption{Client data features.}
\label{tb.features}
\end{table}

We note that one of the limitations of our dataset is that it is biased towards clients who interact with emergency shelters.  An important cohort of individuals experiencing chronic homelessness are the ``rough sleepers'' who prefer to sleep outdoors rather than in shelter.  As we will see in Section~\ref{sec:results}, our algorithms rely heavily on the number of shelter stays in a client's record with predicting chronic shelter use.  A rough sleeper who satisfies the chronic homelessness definition may be overlooked due to being under-represented in the emergency shelter data.  A way to mitigate this bias would be for outreach teams to survey the rough sleeper population and include them in the shelter database.

\section{Prediction Techniques}
\label{sec:alg}

Logistic regression is selected as our first algorithm due to its computational simplicity \parencite{hastie-t-2017} and as a useful benchmark due to its widespread use in the social sector \parencite{diguiseppi-gt-2020, chan-h-2017, hong-b-2018, gao-y-2017, byrne-t-2019}.  Since it is often a challenge for a linear algorithm to converge when operating on multiple features, the logistic regression algorithm utilizes the Sleep and Age data features only.  Both features are normalized to have zero mean and unit variance prior to training and testing.  No regularization algorithm is used on the logistic regression coefficients.

Our second algorithm is a multi-layer perceptron neural network classifier \parencite{goodfellow-i-2016} that operates on all the data features listed in Table~\ref{tb.features}.  The settings for the network are selected using a grid search where predictor accuracy is used as the metric to choose the best setting combination.  The rectified linear function was the best performing hidden layer activation function when combined with an L2 regularization penalty of 0.05, 100 hidden layers and a stochastic gradient descent solver with an adaptive learning rate.  Note that this is a relatively standard neural network design.  There are many additional ways to optimize neural network performance \parencite{goodfellow-i-2016} that we do not explore since the focus of this paper is algorithm evaluation rather than implementation.  

The final algorithm is a simple threshold test that indicates a client will likely become chronic if they sleep in shelter for more than 75\% of the time in the 90 day window.  This corresponds to 67 or more 24 hour periods where a client has accessed sleep services at least once.

\section{Results}
\label{sec:results}

The algorithms described in Section~\ref{sec:alg} are evaluated using k-fold stratified cross-validation \parencite{diamantidis-na-2000}.  Cross-validation is a technique where the dataset is divided into a training set and a testing set.  The machine learning algorithm is trained on the training set but its accuracy is evaluated on the testing set.  Evaluating model accuracy on data excluded from the training prevents the model from over-fitting the data and generating overly optimistic results \parencite{hastie-t-2017}.  The data is randomly divided into 10 equal sized groups (10 {\em folds}) where 9 folds are used for training and 1 for testing \parencite{valmarska-a-2017}.  The algorithms are then evaluated on all 10 possible training/testing fold combinations.  Note that the threshold test described in Section~\ref{sec:alg} does not require training.  When evaluating the threshold test as part of the k-fold cycles, it is tested on the fold designated for testing and the training data is ignored.  

Section~\ref{ssec:metric} presents a comparison of the three predictive techniques using classification metrics.  Section~\ref{ssec:demo} compares the algorithms based on a statistical analysis of the shelter access histories of the cohorts identified by each algorithm.

\subsection{Classification Metrics}
\label{ssec:metric}

Let $P$ denote the 1,549 clients in the dataset flagged as chronic by the Canadian federal definition and $N$ denote the 18,398 - 1,549 = 16,849 clients who are not.  Let the $\hat{\cdot}$ operator denote the client group that a test predicts will become chronic.  Using this notation, the true positives (chronic clients that an algorithm correctly predicts will become chronic) are denoted $\hat{P}$ and the false positives (non-chronic clients that an algorithm incorrectly predicts will become chronic) are denoted $\hat{N}$.

The algorithms are compared using five classification metrics calculated when they are applied to the k-fold cross-validation testing data.  True positive rate or sensitivity ($\hat{P}/P$) is the proportion of chronic clients that we identify correctly.  High sensitivity is important since we do not want to miss anyone in need of help.  False positive rate or false alarm rate ($\hat{N}/N$) is the proportion of non-chronic clients identified as chronic.  A high false positive rate will reduce the efficiency of a housing program since false positives result in referring individuals for support that they may not need.  Confidence or precision is the proportion of clients identified by the algorithm who actually become chronic ($\hat{P}/(\hat{N}+\hat{P})$).  Since $\hat{N}+\hat{P}$ represents the total number of housing referrals, precision is a useful measure since it reflects the fraction of these referrals who would actually become chronic.  Finally, accuracy is defined as the number of correct classifications (true positives plus true negatives) divided by the total population size $(\hat{P} + (N-\hat{N}))/(P+N)$.  The classification metric values for our three classifiers are summarized in Table~\ref{tb.eval}, where the absolute number of true positive and false positives are provided in brackets.  

The results in Table~\ref{tb.eval} suggest that machine learning will lead to the most efficient housing program.  Efficiency is defined as directing resources towards those who need them most \parencite{shinn-m-2019} and Table~\ref{tb.eval} indicates that machine learning algorithms are the best choice for achieving this goal.  For both logistic regression and the neural network, approximately 64\% of clients recommended for housing actually become chronic in the dataset compared to 60\% for the threshold test.  In terms of absolute numbers, the threshold test will refer almost 1000 non-chronic clients for housing while the machine learning algorithms keep this number below 300.  The true positive rates reveal that the threshold test does do a better job of referring a higher proportion of chronic clients for housing. This is because the threshold test essentially casts a much wider net.  It selects a higher number of true positives but also a higher number of false positives.

Table~\ref{tb.eval} includes accuracy since it is a commonly cited performance metric.  However, we will not use it as a means to compare algorithms for this application.  Accuracy can be very misleading when dealing with an imbalanced data set where the condition we are predicting only occurs in a minority of cases.  For example, we could devise a prediction algorithm that simply says no client will ever become chronic.  Since $N$ = 16,849 and $P$ = 1,549, this test would result in 0 true positives and 16,849 true negatives.  The result is an accuracy of $(0+16,849)/18,398$ = 91.6\%.  So, accuracy can sometimes lend credibility to an algorithm that has little or no practical use.

While the focus of this paper is algorithm evaluation and not algorithm design, we can use the results in Table~\ref{tb.eval} to make some comments on machine learning algorithm behavior.  The table shows that logistic regression using two data features performs almost identically to the neural network which uses all ten features.  This indicates that Age and Sleeps are the features with the strongest correlation to chronic shelter use and that the more general neural network algorithm converges to the same linear separation of the data feature space as logistic regression.  It is also instructive to compare the results in Table~\ref{tb.eval} to \parencite{vanberlo-b-2020}.  The authors achieve a very similar confidence performance but a superior true positive rate.  This is most likely due to their use of dynamic time data in addition to summary data as well as a prediction time window that potentially allows their algorithm to use a longer data time window than 90 days.

We should note that any machine learning algorithm tends to find data features that are correlated with an outcome of interest and that correlation does not also imply causation.  The causes of chronic homelessness are many and varied.  Even though our analysis indicates that a larger number of shelter sleeps in the first 90 days is correlated with chronic homelessness, it would be circular to argue that many shelter stays is a cause of chronic homelessness which is defined by a large number of shelter stays.  The data feature correlations we have identified are useful for our application of using emergency shelter data to predict an outcome.  However, taking the additional step of speculating whether these correlations represent some of the underlying causes of chronic homelessness is beyond the scope of this paper.

\begin{table}[htbp]
\centering
\begin{tabular}{c|c|c|c|c|c}
& True Pos.~Rate & False Pos.~Rate  & &  \\
& (Sensitivity) & (False Alarm Rate) & Confidence & Accuracy \\ \hline
Logistic Regression & 0.316 (490) & 0.016 (270) & 0.645 & 0.928 \\
Neural Network & 0.352 (546) & 0.018 (296) & 0.648 & 0.929 \\
Threshold & 0.985 (1526) & 0.058 (982) & 0.608 & 0.945 \\
\end{tabular}
\caption{Binary classification metrics.}
\label{tb.eval}
\end{table}

\subsection{Shelter Access History Comparison}
\label{ssec:demo}

In this section, the three binary classifiers are compared based on the shelter access histories of the cohorts identified by each classifier as chronic.  Each cohort is the union of an algorithm's false positives and true positives (ie. $\hat{N}+\hat{P}$).  In the following, a {\em stay} is a 24 hour period where sleep services are accessed at least once.  An {\em episode} of shelter use is a series of stays where the separation between consecutive stay dates is less than 30 days \parencite{byrne-t-2015}.  Statistical results are presented for the total shelter stays per client, total episodes per client, shelter tenure per client (number of days between first and last stay), shelter use percentage (stays divided by tenure) and the average gap length between episodes of shelter use.  Since shelter client characteristics are often exponentially distributed, we show median, 10th percentile and 90th percentile in addition to average.  The results for logistic regression, the neural network and the threshold test are provided in Tables~\ref{tb.lrDemo}, \ref{tb.nnDemo} and \ref{tb.thshDemo}, respectively.

\begin{table}[htbp]
\centering
\begin{tabular}{c|c|c|c|c}
& Average & Median & 10th Pctl. & 90th Pctl. \\ \hline
Total Stays & 672.7 & 412.5 & 113.0 & 1691.0\\
Total Episodes & 3.8 & 3.0 & 1.0 & 8.0\\
Tenure (days) & 1273.1 & 1055.0 & 199.0 & 2662.0\\
Usage Percentage & 60.8 & 60.8 & 13.7 & 100.9\\
Average Gap Length (days) & 3.1 & 1.5 & 0.9 & 7.3\\ \hline
\multicolumn{5}{c}{Group Size: 760/18398 (4.1\%)}
\end{tabular}
\caption{Logistic regression cohort characteristics.}
\label{tb.lrDemo}
\end{table}

\begin{table}[htbp]
\centering
\begin{tabular}{c|c|c|c|c}
& Average & Median & 10th Pctl. & 90th Pctl. \\ \hline
Total Stays & 661.8 & 394.0 & 108.0 & 1652.0\\
Total Episodes & 3.7 & 3.0 & 1.0 & 8.0\\
Tenure (days) & 1289.0 & 1091.0 & 175.0 & 2677.0\\
Usage Percentage & 59.6 & 59.9 & 13.0 & 100.0\\
Average Gap Length (days) & 3.3 & 1.6 & 1.0 & 7.7\\ \hline
\multicolumn{5}{c}{Group Size: 842/18398 (4.6\%)}
\end{tabular}
\caption{Neural network cohort characteristics.}
\label{tb.nnDemo}
\end{table}

\begin{table}[htbp]
\centering
\begin{tabular}{c|c|c|c|c}
& Average & Median & 10th Pctl. & 90th Pctl. \\ \hline
Total Stays & 563.3 & 362.0 & 120.0 & 1271.0\\
Total Episodes & 5.4 & 4.0 & 1.0 & 11.0\\
Tenure (days) & 1526.9 & 1397.0 & 288.0 & 2961.0\\
Usage Percentage & 44.6 & 36.7 & 11.1 & 93.8\\
Average Gap Length (days) & 4.2 & 2.7 & 1.0 & 9.0\\ \hline
\multicolumn{5}{c}{Group Size: 2508/18398 (13.6\%)}
\end{tabular}
\caption{Threshold test group cohort characteristics.}
\label{tb.thshDemo}
\end{table}

These results show that all three classifiers select groups of clients who would be considered high priority for housing support.  The implication of these results is that the false positives in Section~\ref{ssec:metric} are mostly people who were ``near misses'' and almost but not quite satisfied the chronic homelessness definition.

While the threshold test appeared to be inferior to the machine learning algorithms in Section~\ref{ssec:metric}, its performance is much stronger when evaluated using clients' shelter access histories.  The median number of stays in the threshold test population is approximately equivalent to the machine learning algorithms and the median shelter tenure is actually significantly longer for the threshold test group.  Usage percentage is lower for the threshold population which is most likely due to including longer tenure clients.  The number of episodes of shelter access and the average length of gaps between episodes of shelter use are also both higher for the threshold group.

This suggests that the non-chronic false positive clients selected by the threshold test exhibit episodic shelter access characteristics.  Originally discovered by the cluster analysis in \parencite{kuhn-r-1998}, episodic shelter clients are characterized by more sporadic periods of shelter access over a long period of time.  While perhaps not strictly ``chronic'', we argue that any client group with 362 median stays and a median period of shelter interaction of 1397 days is clearly in need of housing support.

One possible drawback of the threshold test is that it identifies a greater number of clients (13.6\%) compared to the machine learning algorithms (4.1\% and 4.6\%).  This may raise concerns that its use may overwhelm a housing program.  To put this in perspective, the 2,508 clients identified by the threshold test are stretched over 101 months of data.  This means the test would identify an average of approximately 24.8 clients per month.  The current housing programs at the Calgary Drop-In Centre have successfully been able to house approximately twice this many clients \parencite{di-community-report-2019}.  Therefore, we believe that the clients identified by the threshold test would fall within the capacity of most housing programs.

\section{Concluding Remarks}
\label{sec:concl}

Data science has the potential to be a very powerful tool for delivering efficient programs to support people experiencing homelessness.  This exciting area of research is a true inter-disciplinary collaboration where ideas and techniques from statistics, engineering and computer science are applied to problems in the social science domain.  However, it is important to evaluate the benefit of these techniques using an approach that recognizes that any group of people will have characteristics that cannot be easily summarized by a binary label like ``chronic'' or ``not chronic''.  This has important implications not just for understanding the nature of a group identified for support but for choosing the technique used to identify that group in the first place.  

For data science to have benefit, it must make the transition from the research domain to application in a front-line shelter or social program setting.  Implementing many machine learning algorithms comes with technical barriers that are not trivial to overcome, particularly in the not-for-profit sector.  If a simple technique offers similar performance to a complex one, the simple technique should be chosen every time.

The broad contribution of this paper is to highlight the importance of including a client shelter access history analysis when using data science techniques to support homelessness services.  By applying this philosophy to the specific problem of predicting chronic homelessness, our second contribution is to show that a simple threshold test offers similar performance to more sophisticated machine learning algorithms.  This test could be implemented today by most shelters and housing programs using IT infrastructure already in place.  This is an example of how proper evaluation of data science can actually remove some of the barriers to using these techniques in a real world setting.

\section{Acknowledgments}
\label{sec:ack}

The authors would like to acknowledge the support of the Natural Sciences and Engineering Research Council of Canada (NSERC), the Calgary Drop-In Centre and the Government of Alberta.  This study is based in part on data provided by Alberta Seniors, Community and Social Services. The interpretation and conclusions contained herein are those of the researchers and do not necessarily represent the views of the Government of Alberta. Neither the Government of Alberta nor Alberta Seniors, Community and Social Services express any opinion in relation to this study.

\printbibliography

\end{document}